\documentclass{aa}
\usepackage{graphicx}
\usepackage{natbib}
\usepackage{txfonts}
\bibpunct{(}{)}{;}{a}{}{,}
\newcommand{\ha}{H$\alpha$~}
\newcommand{\kms}{\,km\,s$^{-1}$}

\newcommand{\cmt}{\,cm$^{-3}$}
\newcommand{\myr}{\,$M_{\sun}\,{\rm yr}^{-1}$}

\newcommand{\ecsa}{$\rm\,erg\,cm^{-2}\,s^{-1}\,\AA^{-1}$}
\begin{document}

\title{Broad H${\alpha}$ wings from the optically thin stellar wind\\ 
       of the hot components in symbiotic binaries}

\author{A.~Skopal \inst{}}

\institute{Astronomical Institute, Slovak Academy of Sciences,
           059\,60 Tatransk\'{a} Lomnica, Slovakia \\
           \email{skopal@ta3.sk}
          }

\date{Received / Accepted }

\abstract
 {}
% Aims
 {To model broad \ha wings observed in symbiotic binaries by 
  an optically thin, bipolar stellar wind from their hot 
  components as an alternative to that considering the Raman 
  scattering of Ly$\beta$ photons on atomic hydrogen.}
% Methods
 {Profile-fitting analysis. Comparison of the observed broad \ha 
  wings and their luminosity with those predicted by the model.} 
% Results
 {Synthetic \ha profiles fit excellently the observed wings for 
  $| \Delta v | \ga $\,200\kms\ in our sample of 10 symbiotic 
  stars during the quiescent as well as active phases. 
  The wing profile formed in the stellar wind can be approximated 
  by a function $f(\Delta v) \propto \Delta v^{-2}$, which is of 
  the same type as that arising from the Raman scattering. 
  Therefore it is not possible to distinguish between these 
  two processes only by modeling the line profile. 
  Some observational characteristics of the H$\alpha$-emission, 
  its relationship with the emission measure of the symbiotic 
  nebula and a steep radio spectrum at 1.4 -- 15\,GHz suggest 
  the ionized stellar wind from the hot component to be 
  the dominant source contributing to the \ha wings during 
  active phases. The model corresponding mass-loss rates from 
  the hot components are of a few $\times 10^{-8}$\myr\ and 
  of a few $\times\,(10^{-7} - 10^{-6})$\myr\ during quiescent 
  and active phases, respectively.}
 {}
\maketitle
%
%----------------------------------------------------------------------
%
\section{Introduction}

\cite{vw+93} and \cite{i+94} presented a large survey of high- 
and low-resolution \ha line profiles of symbiotic stars. 
The profiles showed broad wings, in most cases extended well 
within the presented wavelength range of 6540 -- 6580\,\AA. 
Their origin has been investigated by number of authors. 
A popular interpretation assumes Raman scattering of Ly$\beta$ 
photons on atomic hydrogen to be responsible for filling in 
the broad \ha wings. This possibility was firstly pointed out 
by \cite{nsv89} and the corresponding quantitative model was 
elaborated by \cite{l00} and \cite{lh00}. 
Other possibilities for the \ha wing formation mechanism -- 
rotating disks, electron scattering, fast stellar wind 
and \ha damping wings -- were also discussed. 
\cite{r+94} modeled the \ha profiles on the assumption 
that they originate in an accretion disk. Acceptable fits 
were found only for CH\,Cyg, AG\,Dra and T\,CrB. 
Generally, the model wings were broader than the observed ones. 
   The possibility of the electron scattering was analyzed by 
\cite{at03} for a representative case of M2-9. They 
found unrealistically high values of the electron temperature 
and concentration for the electron-scattering region. 
   Concerning to the \ha damping wings there is no 
elaborated application for symbiotic binaries. \cite{l00} 
only discussed briefly this possibility for the case of SY\,Mus 
\citep{schmutz}. He came to conclusion that the wing emission 
arises in a much more extended region then that producing 
the line core. 
\cite{sk+02} modeled the extended \ha wings from 
active phases of CH\,Cyg by a spherically symmetric and 
optically thin stellar wind. A comparison between the modeled 
and observed profiles was satisfactory and also the derived 
mass-loss rate was in agreement with that suggested by the radio 
observations. Therefore we propose the fast stellar wind from 
the hot component in symbiotic binaries to be the most 
promising alternative to that considering the Raman scattering 
process. 

Accordingly, in Sect.~2 we introduce a model of a bipolar 
stellar wind at the optically thin limit to calculate the 
broad \ha wings. 
In Sect.~3 we compare our model profiles with those observed 
during quiescent and active phases of selected symbiotic stars. 
In Sect.~4 we discuss observational characteristics of \ha 
profiles connected with the hot star wind. 
%
%==============================================================|
%-- Fig. 1.: A schematic picture of the stellar wind model  ---|
%==============================================================|
%
\begin{figure*}
\centering
\begin{center}
\resizebox{\hsize}{!}{\includegraphics{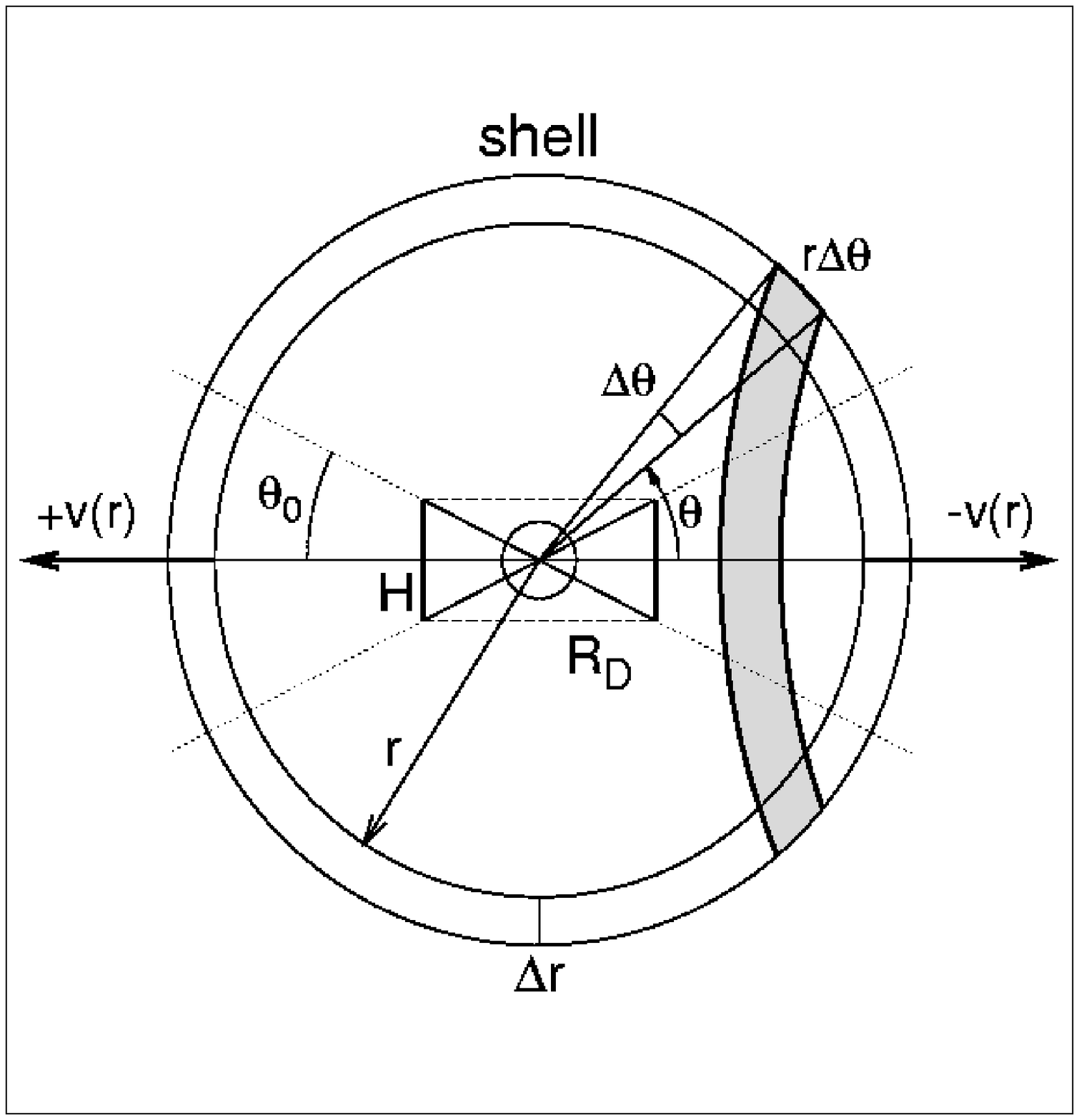}\hspace*{0mm}
                      \includegraphics{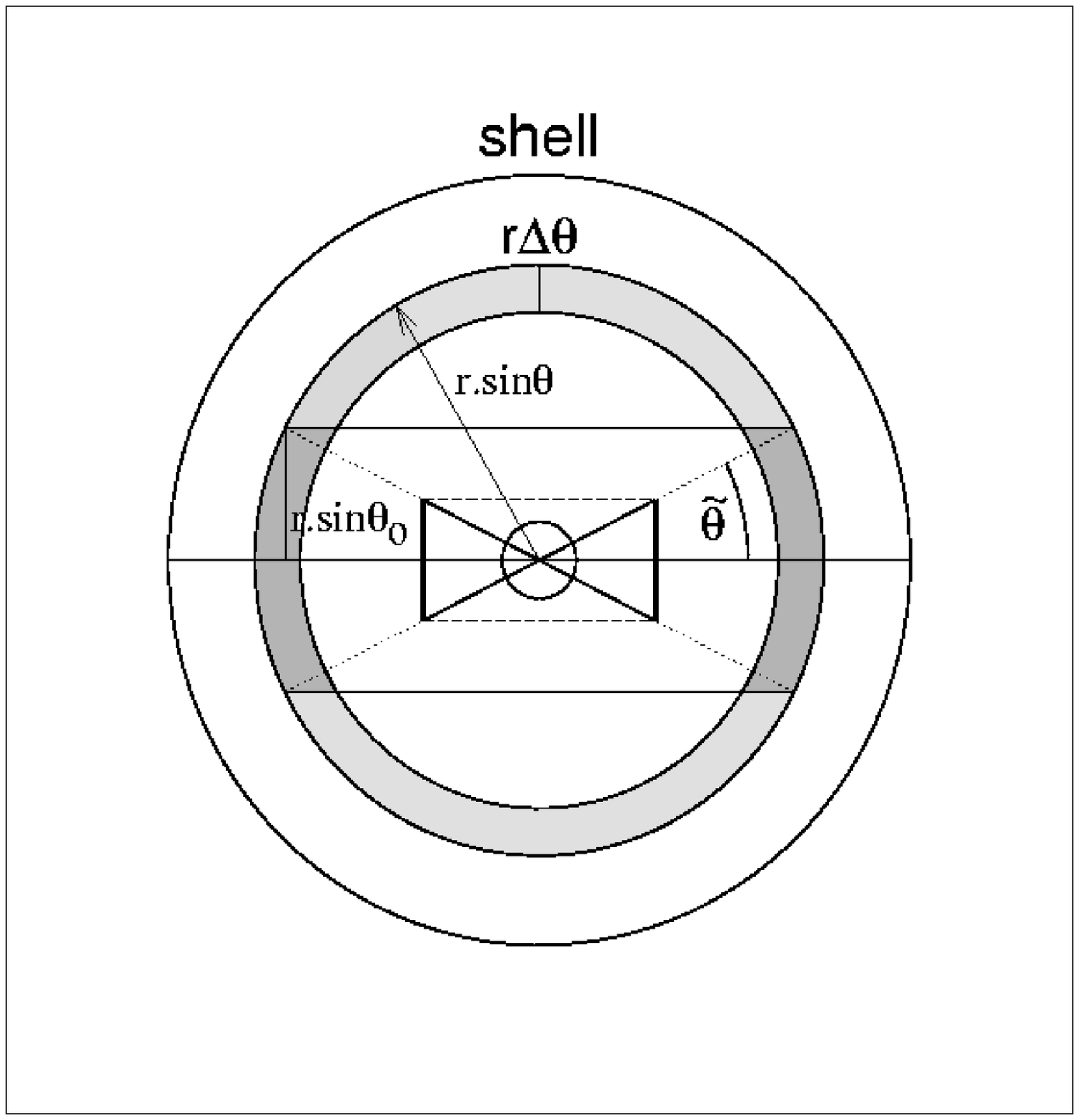}}
\caption[]{
Geometry of the wind model we used to calculate the line 
profile of \ha and its luminosity. The wind is produced by 
the central star (small circle at the mid of panels). 
A fraction of the wind is blocked by an optically thick 
disk/torus at the center that cut out a cone from the sphere 
with the opening angle $\pi - 2\theta_0$. By this way we 
simulate bipolar geometry of the stellar wind. 
The shadow belt on the side view (left) and/or the annulus from 
the frontal view (right) represents a part of the shell with 
the same radial velocity, $-v(r)\cos(\theta)$. Radiation from 
the annulus, cut out by 4$\tilde\theta$ radians, is blocked 
by the central disk (the darker part). The angle $\tilde\theta$ 
is counted in the plane containing the annulus. 
          }
\end{center}
%\label{fig_1}
\end{figure*}

\section{Model of the bipolar wind}

\subsection{Signatures of the mass outflow}

Here we summarize main observational features of a mass 
outflow indicated for active symbiotic stars. They are: 
  (i) 
Broadening of emission line profiles and/or the P-Cygni 
type of profiles represent direct indications of 
a mass-flow from the star. Typical velocities are a few 
hundred of \kms\ \citep[e.g.][]{fc+95,nsv95,sk+97}. 
  (ii)
A significant enhancement of the nebular emission in the 
continuum by a factor of $\approx$10 relatively to quiescent 
phases \citep[][ Tables~3 and 4]{sk05} can in part result from 
a supplement of new emitters into the particle bounded nebula, 
for example, due to an increase in the mass-loss rate from 
the active object. 
  (iii)
The radio light curves usually show a decline at beginnings 
of outbursts with an increase after the optical maximum 
\citep[e.g.][]{fc+95,b+04}. In some cases evolution from 
a point source to a spatial structure was observed 
\citep[e.g. CH\,Cyg and Z\,And:][]{kenny,b+04}. This can be 
a result of a high-velocity mass-outflow, which gradually 
becomes optically thin at radio wavelengths. Velocities from 
a few hundred of \kms\ to 1000$\div$2000\,\kms\ can be derived 
from images \citep[e.g. CH\,Cyg,][]{crok+01,crok+02}. 
  (iv) 
The $X$-ray emission can be also explained by interaction of 
the outflowing material with surrounding nebular gas. 
The extended $X$-ray emission in the CH\,Cyg $Chandra$ image 
was aligned with the optical and radio jets \citep{g+s04}. 
Thus the velocities connected with the $X$-ray emission 
can be similar to those derived from the radio. 
For Z\,And, \cite{sok+06} ascribed the $X$-ray emission from 
its major 2000-03 outburst to the shock-heated plasma as 
a consequence of the mass ejection from the white dwarf. 
  (v)
Emission line profiles of forbidden lines from highly ionized 
atoms can be produced by the wind-wind collision zone in a 
binary system \citep[e.g.][]{w+84,ejw}. 
Aspects of the wind-dynamics including colliding winds in 
symbiotic binaries were reviewed by \cite{w+f00}. 

Finally, we note that different features of the outflowing 
material in the spectrum can reflect different driving 
mechanisms. For example, ejected 
rings or shells produce broad profiles with rather steep sides 
of all lines (FWHM\,$\sim0.7\times$\,FWZI). Classical novae 
1494\,Aql and V475\,Sct demonstrate this case 
\citep[][]{eyres+05,chochol+05}. 
For a star with a spherically symmetric and optically thin 
wind, the line emissivity is proportional to the square of 
the particle concentration, which is diluted with the radial 
distance $r$ as $1/r^2$. In this case a strong line core, 
originating from the vicinity of the wind source, is accompanied 
with faint extended wings from large distances (FWHM\,$\ll$\,FWZI). 
In the following model we consider this case. 

\subsection{Geometry of the wind}

The geometry of our wind model is introduced in Fig.~1. 
The model assumes an optically thin stellar wind with a 
spherically symmetric structure and the origin at/around 
the central star. 
According to \cite{sk05} we put an optically thick 
disk/torus, characterized with the height $H$ and radius 
$R_{\rm D}$, to the center of the hot object. We assume 
the disk to be seen edge-on due to a high orbital inclination. 
The outer rim of the disk cut out the angle 
   2$\theta_0 = 2\tan^{-1}(H/R_{\rm D})$ 
from a sphere with the center at the accretor, and by
this way simulates bipolar shape of the stellar wind with
the opening angle $\pi - 2\theta_0$ radians (Fig.~1 left).
The wind with this geometry produces line profiles that are 
symmetrical with respect to the reference wavelength 
($\lambda_0$). 
We assume that the particle concentration $n(r)$ at any point 
in the wind is related to the mass loss rate $\dot M$ and 
the velocity $v(r)$ via the mass continuity equation, i.e. 
%
%----------------------- Eq. 1 ----------------------
\begin{equation}
 n(r) = \dot M/4\pi r^{2}\mu m_{\rm H} v(r),
\end{equation}
%----------------------------------------------------
where $\mu$ is the mean molecular weight and $m_{\rm H}$ is 
the mass of the hydrogen atom. 
According to the \cite{cak} model we approximate the velocity 
distribution in the hot star wind by 
%
%----------------------- Eq. 2 ----------------------
\begin{equation}
 v(r) = v_{\infty}(1 - R_{\rm w}/r)^{\beta}. 
\end{equation}
%----------------------------------------------------
The velocity $v(r)$ of the wind increases monotonically outward 
from its beginning at $R_{\rm w}$ and asymptotically approaches 
the terminal speed $v_{\infty}$. 
The distance $r$ is counted from the center of the star and
the parameter $\beta$ characterizes an acceleration of the
wind, i.e. the 'slope' of $v(r)$. A smaller $\beta$ 
corresponds to a faster transition to $v_{\infty}$ 
\citep[e.g.][]{b+85}. 

\subsection{\ha luminosity}

In our simplified approach we assume that the wind is fully 
ionized and completely optically thin in \ha with a steady 
mass-loss rate. 
The optically thin case is supported by the large velocity 
gradient in the stellar wind, because of its large terminal
velocity ($v_{\infty} \gg v_{\rm th}$, where $v_{\rm th}$ is
the thermal velocity). If a line photon, created by 
recombination in such a wind, has traveled a distance 
$l > 2 v_{\rm th}/(dv/dl)$, it is Doppler shifted with 
respect to the surrounding gas by more than 2$v_{\rm th}$ and 
thus cannot be absorbed any more in the same line transition 
\citep[e.g.][]{lc99}. 
Under such conditions the escape probability of the emitted 
photons will be close to 1. However the optically thin condition 
can be attained only at large distances from the source of 
the wind. 
A good agreement between the observed and modeled profiles 
for $|\Delta v| \ga $\,200\kms\ (see below, Fig.~3) suggests 
the validity of the optically thin regime from about 
1.2$\div$1.5\,$R_{\rm w}$ (Eq.~(2) and parameters from Table~1). 

The total line luminosity, $L(\rm H\alpha$), is related 
to the line emissivity of the wind, 
$\varepsilon_{\alpha} n_{\rm e}n^{+}$, by
%
%----------------------- Eq. 3 ----------------------
\begin{equation}
  L({\rm H}\alpha) = 
    4\pi\varepsilon_{\alpha}\!\! \int_{r_{\rm i}}^{\infty}\!\!
                     n_{\rm e}n^{+}(r)[1-w(r)]\,r^2{\rm d}r,
\end{equation}
%----------------------------------------------------
where $r_{\rm i}$ is a certain distance from the source of 
the wind where the integration starts from (its quantity 
is given by the model; $r_{\rm i} \ge H$, $R_{\rm w} < H$), 
$\varepsilon_{\alpha} = 
   3.56\times 10^{-25}\rm erg\,cm^{3}\,s^{-1}$ 
is the volume emission coefficient in \ha for 
$T_{\rm e}$ = 10$^{4}$\,K. We assume it to be constant 
throughout the wind. $n_{\rm e}$ and $n^{+}$ are concentrations 
of electrons and ions (protons). We assume a completely ionized 
medium ($n_{\rm e} \simeq n^{+}$) and radial distribution 
of particles as given by Eq~(1). 
The factor $w(r)$ determines visibility of the wind for 
the outer observer. It can be expressed as 
%
%%----------------------- Eq. 4 ----------------------
\begin{eqnarray}
  w(r) = \sin\theta \, = \, H/r ~~~~~~ {\rm for}~~~ r < r_0, 
\nonumber \\
       = \sin\theta_0 \, = \, H/r_0 ~~~ {\rm for}~~~ r > r_0 ,
\end{eqnarray}
where $r_0^2 = R_{\rm D}^2 + H^2$ (Fig.~1). 
Therefore we have to integrate contributions from the shells 
between $H$ and $r_0$ 
%(i.e. ${\rm d} V = 4\pi r^2 (1 - H/r) {\rm d}r$) 
and those above the $r_0$ radius 
%(i.e. ${\rm d}V = 4\pi r^2 (1 - H/r_0) {\rm d}r$)
separately. 
Substitution of Eqs.~(1), (2) and (4) into Eq.~(3) and using 
dimensionless parameters 
  $x=R_{\rm w}/r$, 
  $\alpha = H/R_{\rm w}$ 
and 
  $f=R_{\rm w}/r_0$ 
yields an expression for the \ha luminosity as
%
%%----------------------- Eq. 5 ----------------------
\begin{equation}
  L({\rm H}\alpha) = \frac{\varepsilon_{\alpha}}{4\pi(\mu m_{\rm H})^2}\,
          \Big(\frac{\dot M}{v_{\infty}}\Big)^{2}\frac{1}{R_{\rm w}}
          \times (I_1 + I_2),
\end{equation}
where
%
%%----------------------- Eq. 6 ----------------------
%\begin{equation}
\begin{eqnarray}
  I_1 = \int_{f}^{1/\alpha}\!
        \frac{1-\alpha x}{(1-x)^{2\beta}}{\rm d}x\, = 
~~~~~~~~~~~~~~~~~~~~~~~~~~~~~~~~~~~~
\nonumber \\
 =\,\frac{\alpha}{2(\beta-1)}\,\Big[\big(1-1/\alpha\big)^{2(1-\beta)} 
                            - \big(1-f\big)^{2(1-\beta)}\Big]
\nonumber \\
  +\, \frac{1-\alpha}{2\beta-1}\,\Big[\big(1-1/\alpha\big)^{1-2\beta}
                            - \big(1-f\big)^{1-2\beta}\Big]
\nonumber \\
{\rm for}~~ \beta \not= 0.5, 1,
\nonumber \\
  =\, \alpha\ln\frac{\alpha(1-f)}{\alpha-1}\,- \frac{1-\alpha f}{1-f} 
~~~~~~~~~{\rm for}~~ \beta = 1,
\nonumber \\
~~~~~  = (1- \alpha)\ln\frac{\alpha(1-f)}{\alpha-1} + 1 - \alpha f 
~~{\rm for}~~ \beta = 0.5 
\end{eqnarray}
%\end{equation}
%
and 
%
%%----------------------- Eq. 7 ----------------------
\begin{equation}
  I_2 = \int_{0}^{f}\!
        \frac{1-H/r_0}{(1-x)^{2\beta}}{\rm d}x 
      = \frac{1-H/r_0}{(1-2\beta)}\,\Big[1-\big(1-f\big)^{1-2\beta}\Big] .
\end{equation}
Analytical expression of the integral $I_2$ was already 
introduced by \cite{sk+02}. 
Thus comparing the \ha luminosity from observations 
to that predicted by Eq.~(5) allow us to estimate 
the mass-loss rate, $\dot M$. 
To determine theoretical values of $L({\rm H}\alpha)$ 
requires a knowledge of parameters characterizing the wind: 
  $v_{\infty}$, 
  $R_{\rm w}$, 
  $\beta$ 
and 
  $\theta_0$. 
To estimate their appropriate quantities we fit 
a synthetic-line profile to the observed one. This requires 
a different manner of integration of the wind's 
contributions. We introduce it in the following section. 
%
%===================================================|
%----- Fig. 2 : La(200) as a function of dM/dt -----|
%===================================================|
%
\begin{figure*}
\centering
\begin{center}
\resizebox{16cm}{!}{\includegraphics[angle=-90]{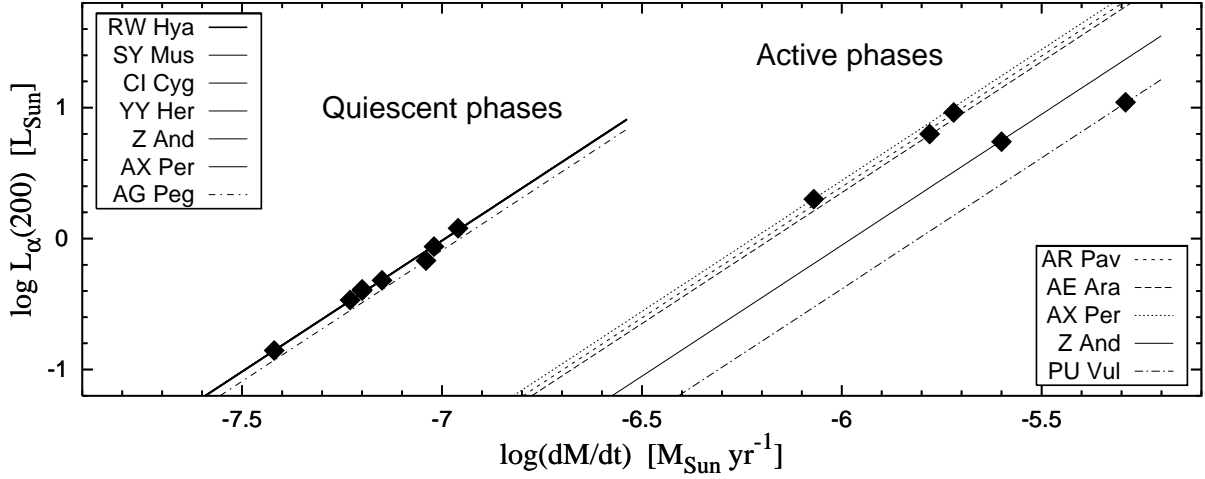}}
\caption[]{
Luminosity of \ha wings for $|\Delta v| \ge 200$\kms, 
$L_{\alpha}$(200), as a function of the mass-loss rate 
calculated according to Eq.~(12) for parameters $R_{\rm w}$, 
$\beta$ and $v_{\infty}$ in Table~1. 
The observed quantities of $L_{\alpha}$(200) are denoted 
by filled diamonds (Table~1). 
          }
\end{center}
\end{figure*}

\subsection{\ha profile}

To reconstruct the global line profile we redistribute
the independent Doppler-shifted contributions from each
volume element of the wind expanding material in the
radial velocity co-ordinates. The profile thus represents
a 'broadening function' resulting from a field of contributions
which differ in emissivity and radial velocity. It can be 
compared only to high-velocity features in the profile 
produced by regions with $\tau({\rm H}\alpha) < 1$. 
A technique of integration can be understand with the help
of Fig.~1. Volume elements of the same radial velocity, 
$RV = -v(r)\cos\theta$, are represented by annuli around 
the line of sight, and can be expressed as 
%
%%----------------------- Eq. 8 ----------------------
\begin{equation}
\Delta V = 
  (2\pi - 4\tilde\theta) r^2 \sin\theta \Delta\theta \Delta r, 
\end{equation}
where the angle 4$\tilde\theta$ corresponds to a fraction of 
the annuli, which radiation is blocked by the central disk. 
The $\tilde\theta$ angle is counted in the plane containing 
the annulus of the radius $r \sin\theta$ and is related to 
$H$ and $\theta$ as (Fig.~1 right) 
%
%%----------------------- Eq. 9 ----------------------
\begin{eqnarray}
  \sin\tilde\theta =
      \frac{H}{r \sin\theta}~~~~~~{\rm for}~~~ H < r < r_0,
\!\!\!\!\!\!\!\!\!\!\!\!
\nonumber \\
        = \frac{\sin\theta_0}{\sin\theta}~~~~~~~{\rm for}~~~ r > r_0. 
\end{eqnarray}
In the sense of Eq.~(3), radiative contributions of such 
cut-out annuli are 
    $\Delta L({\rm r},\theta) = 
        \varepsilon_{\alpha} n({\rm r})^2 \Delta V$. 
They correspond to a certain radial velocity $\Delta v$ 
(or $\Delta \lambda$) in the profile. With the aid of 
Eqs.~(1) and (8) we can write them as 
%
%%----------------------- Eq. 10 ----------------------
\begin{equation}
 \Delta L(r,\theta) =
 \xi \frac{(2\pi - 4\tilde\theta)\sin\theta}
          {r^2 {\rm v}(r)^2}\Delta\theta \Delta r ~~~ [{\rm erg\,s^{-1}}],
\end{equation}
where ${\rm v}(r) = v(r)/v_{\infty}$ and the factor 
%
%%----------------------- Eq. 11 ----------------------
\begin{equation}
 \xi = \frac{\varepsilon_{\alpha}}{(4\pi\mu m_{\rm H})^2}
       \Big(\frac{\dot M}{v_{\infty}}\Big)^2.
\end{equation}
Redistributing all the visible emissions according to their 
radial velocities we obtain the resulting line profile. 
Integration of each shell begins from the direction at 
$\theta = \theta_1$ and ends at $\theta = \pi - \theta_1$. 
Their contributions are summarized from $H$ to a distance at 
which the wind's emission can be neglected. Thus the line 
luminosity can be expressed as 
%
%%----------------------- Eq. 12 ----------------------
\begin{equation}
 L({\rm H}\alpha) = \xi \int_{H}^{\infty}\!
          \int_{\theta_1}^{\pi - \theta_1}\! 
          \frac{(2\pi - 4\tilde\theta)\sin\theta}
               {r^2 {\rm v}(r)^2}
                          {\rm d}\theta{\rm d}r. 
\end{equation}
According to the relation (9), integration has to be divided 
into two parts: for $H < r < r_0$, $\theta_1 = \sin^{-1}(H/r)$ 
and for $r > r_0$, $\theta_1 = \theta_0$. 
Contributions from the nearest regions to the wind origin, 
$R_{\rm w}$, are characterized with $H < r \ll r_0$ and 
large values of $\theta_1$. According to relations (1) and 
(2) they are strong due to a high density of the wind and 
contribute mainly to the line center, because of a small 
radial velocity. However, amount of their emission is reduced 
significantly by the large value of $\sin\tilde\theta$ in 
our model (Eq.~9), i.e. a significant fraction of radiation 
is blocked by the torus. 
Integration ends at a finite limit of 200\,$R_{\odot}$ 
\citep[Fig.~C1 in][]{sk+02}. After summing of all the 
contributions we scaled the synthetic profile to the maximum 
of the observed one. Examples are shown in Fig.~3. 
The input parameters, $v_{\infty}$ and $\theta_0$ can be 
inferred from observations, while the resulting parameters, 
$R_{\rm w}$, $\xi$ and $\beta$, are given by an appropriate 
fit to the observed profile (Sect.~3.2). 
If we rewrite the relation (12) in the form 
%
%%----------------------- Eq. 13 ----------------------
\begin{equation}
 L({\rm H}\alpha) = \xi \times I_3  = 4\pi d^{2} F({\rm H}\alpha),
\end{equation}
we can express the ratio 
%
%%----------------------- Eq. 14 ----------------------
\begin{equation}
  \frac{\dot M}{v_{\infty}} = 8.5 \times 10^{-10}\,
                              \Big(
           \frac{F({\rm H}\alpha)}{I_3}
                              \Big)^{1/2}\!\times\,d ~~~~~~~~ 
           \frac{M_{\sun}\,{\rm yr}^{-1}}{\rm km\,s^{-1}} ,
\end{equation}
where the distance $d$ is in kpc and the observed bolometric 
flux $F({\rm H}\alpha)$ in $\rm erg\,cm^{-2}\,s^{-1}$. 
Following to Eq.~(10) integral $I_3$ is given by the geometry 
of the wind and is proportional to the sum of all its visible 
volume elements. For wing fluxes, e.g. $F_{\alpha}$(200) 
(see Sect.~3.3), the $I_3$ integral includes contributions 
only with the radial velocity $|\Delta v| > $\,200\kms. 
%
%==================================================|
%-------------- Table 1: Mass-loss ----------------|
%==================================================|
%
\begin{table*}
\caption[]{
Parameters of \ha models ($R_{\rm w}$, $\beta$, $v_{\infty}$), 
observed luminosities and corresponding mass-loss 
rates ($\dot M$) and emission measures ($EM_{\rm w}$) of the 
hot star wind. $L_{\alpha}$(200) denotes the luminosity of the 
\ha wings for $|\Delta v| \ge 200$\kms, while $L_{\alpha}(0)$ 
is the total line luminosity. Distances are from \cite{sk05}.
          }
\begin{center}
\begin{tabular}{cccccccccc}
\hline
\hline
 Object          &
   d             &
  Date           &
$R_{\rm w}$      &
$\beta$          &
$v_{\infty}$     &
$L_{\alpha}(0)$  &
$L_{\alpha}$(200)& 
log($\dot M$)    &
$EM_{\rm w}$      \\
                 &
[kpc]            &
                 &
[$R_{\sun}$]     &  
                 &
[km\,s$^{-1}$]   &
[$L_{\sun}$]     &
[$L_{\sun}$]     &
[$M_{\sun}\,\rm yr^{-1}$] &
[cm$^{-3}$]      \\
%% Obj.    d     Date    Rw   beta  v(oo)   La   L(200)   dM/dt   EM_w
\hline\\[-3mm]
\multicolumn{9}{c}{Quiescent phases}\\[1mm]
%
% $R_{\rm D} = 0.5\,L_{\sun}$, $H = 0.05\,L_{\sun}$}   \\[1mm]
%
RW\,Hya & 0.8 & 10/07/92 & 0.040 & 1.70 &1000& 1.1 & 0.34 &-7.23&2.0
              $\times 10^{58}$\\
 Z\,And & 1.5 & 22/09/88 & 0.040 & 1.72 &1000& 5.1 & 0.87 &-7.02&5.4
              $\times 10^{58}$ \\
AX\,Per & 1.7 & 01/11/93 & 0.042 & 1.70 &1000& 6.6 & 1.2  &-6.96&9.2
              $\times 10^{58}$ \\
CI\,Cyg & 2.0 & 21/09/88 & 0.042 & 1.72 &1000& 8.0 & 0.40 &-7.20&3.2
              $\times 10^{58}$ \\
        &     & 20/06/89 & 0.040 & 1.63 &1000& 3.3 & 0.48 &-7.15&2.6
              $\times 10^{58}$ \\
AG\,Peg & 0.8 & 15/07/88 & 0.042 & 1.70 &1200& 4.6 & 0.68 &-7.04&4.4
              $\times 10^{58}$ \\
YY\,Her & 6.3 & 20/06/89 & 0.040 & 1.75 &1000& 2.8 & 0.40 &-7.20&2.5
              $\times 10^{58}$ \\
SY\,Mus & 1.0 & 14/07/88 & 0.042 & 1.70 &1000& 3.0 & 0.14 &-7.42&1.1
              $\times 10^{58}$ \\[1mm]
\multicolumn{9}{c}{Active phases}\\[1mm]
%
% $R_{\rm D} = 3.5\,L_{\sun}$, $H = 1.05\,L_{\sun}$}\\[1mm]
%
AR\,Pav & 4.9 & 14/07/88 & 0.90 & 1.75 &1800& 39.5& 9.2 & -5.72
              &5.2$\times 10^{59}$ \\ 
AE\,Ara & 3.5 & 14/07/88 & 0.90 & 1.80 &2000& 36.4& 6.3 & -5.78
              &3.6$\times 10^{59}$ \\
AX\,Per & 1.7 & 22/09/88 & 0.85 & 1.80 &1600&  7.1& 2.0 & -6.07
              &9.8$\times 10^{58}$ \\
Z\,And  & 1.5 & 11/12/00 & 1.70 & 1.70 &2600& 20.8& 5.5 & -5.60
              &2.9$\times 10^{59}$ \\
PU\,Vul$^{\ast}$ & 3.2 & 28/09/88 & 1.67   & 1.70 &2100& 144 & 
11$^\dagger$  & -5.29 &1.5$\times 10^{60}$  \\
\hline
\end{tabular}
\end{center}
$^{\ast}$ $d = 3.2$\,kpc, $E_{\rm B-V}$ = 0.22 \citep{rudy},~~
$^\dagger$ = $L_{\alpha}$(500)
%$^\ddagger$ = $\dot M$(500)
\end{table*}
%
%
%====================================================================|
%-- Fig. 3: H-alpha profiles from Q and A: observations + models  ---|
%====================================================================|
%
%
\begin{figure*}
\centering
\begin{center}
            {\sf Q U I E S C E N T ~ P H A S E S}\\[-1mm]
\resizebox{\hsize}{!}{\includegraphics[angle=-90]{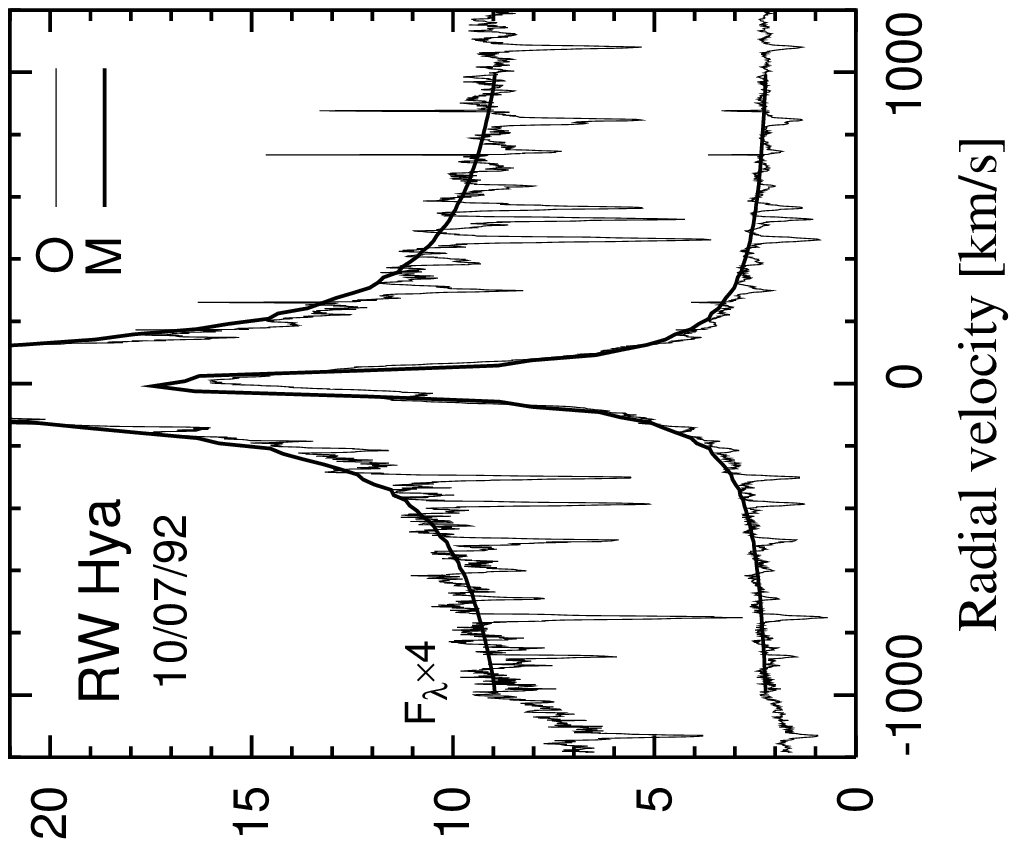}
                      \includegraphics[angle=-90]{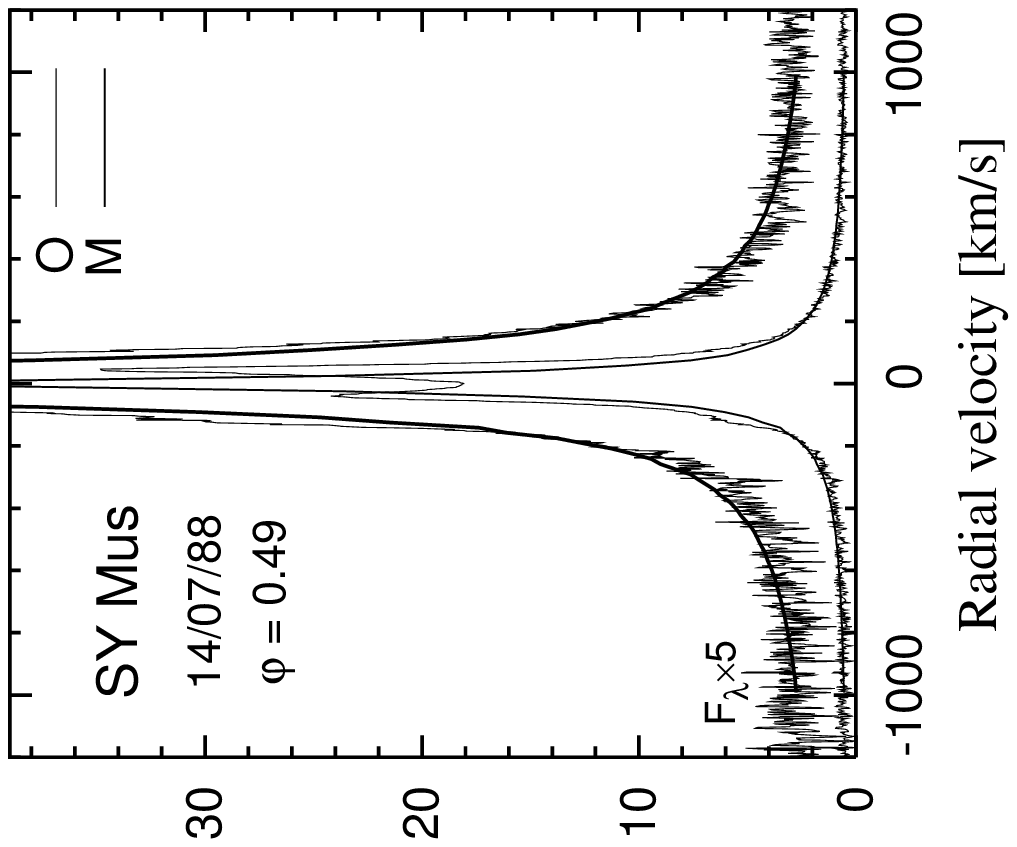}
                      \includegraphics[angle=-90]{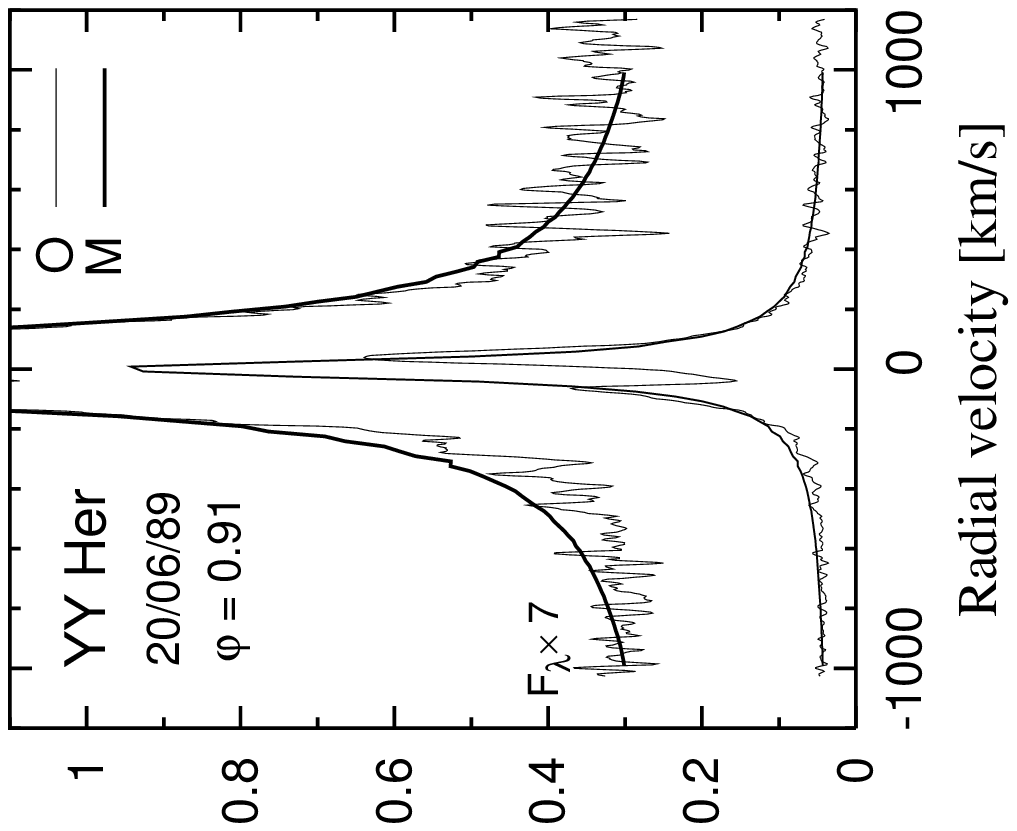}
                      \includegraphics[angle=-90]{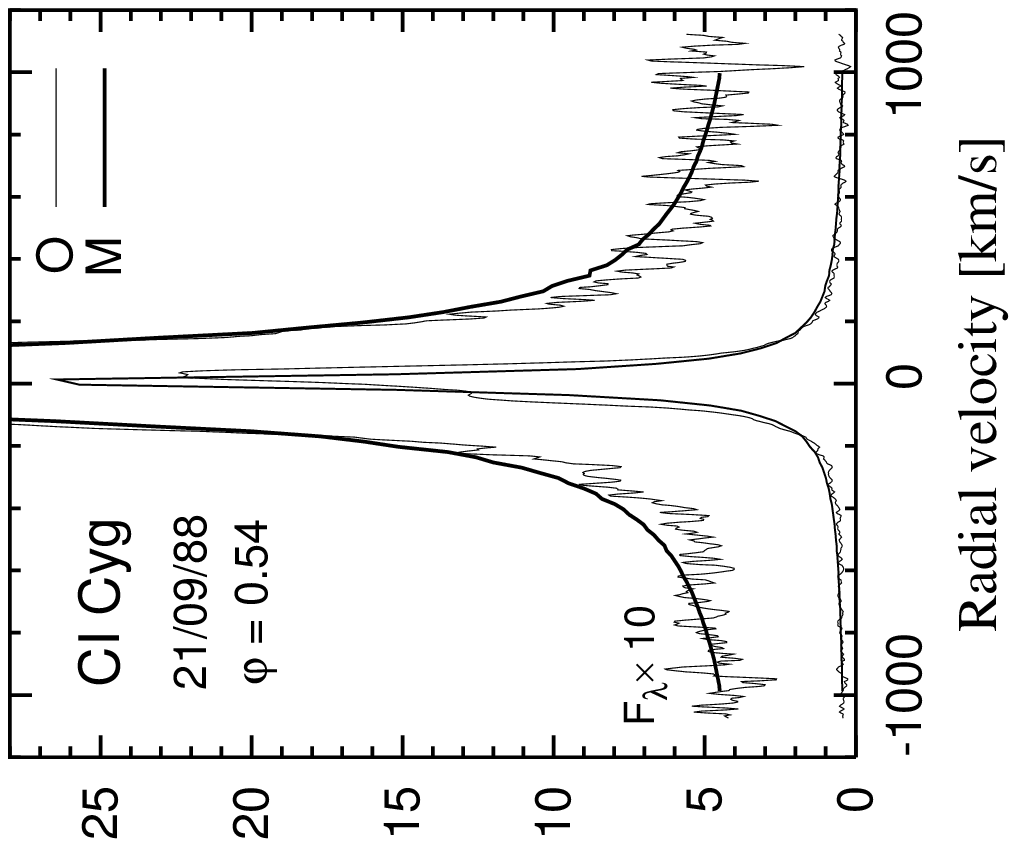}}
\resizebox{\hsize}{!}{\includegraphics[angle=-90]{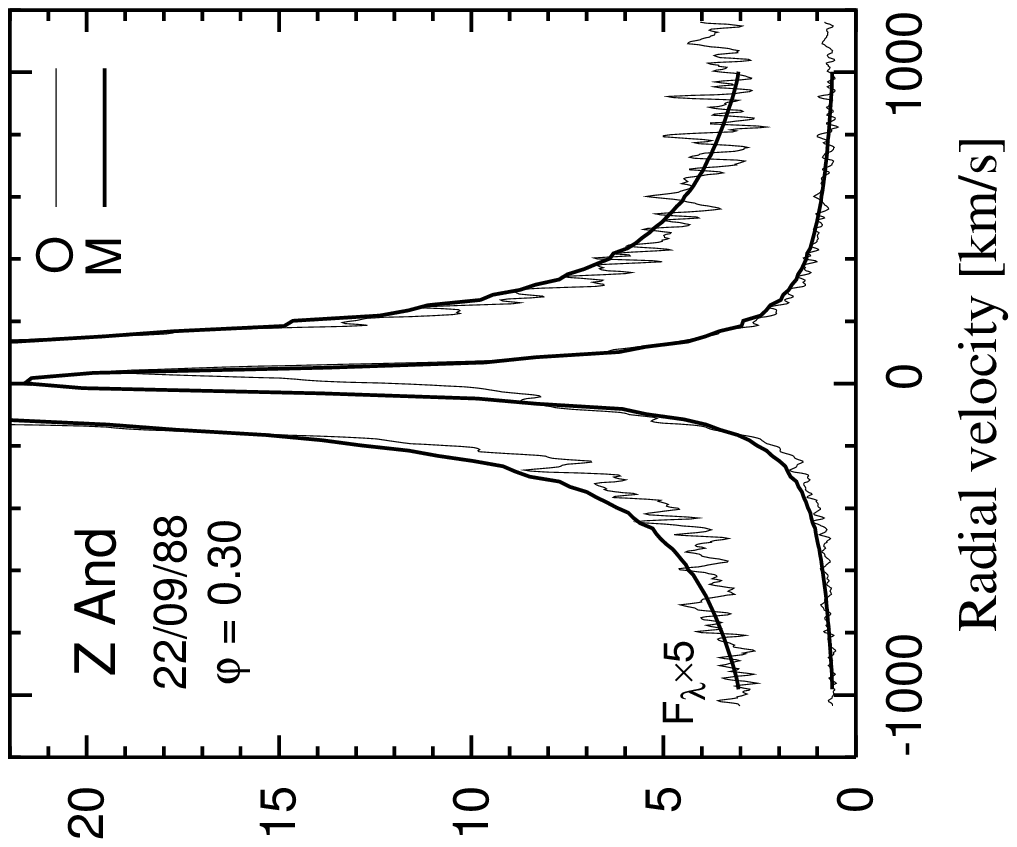}
                      \includegraphics[angle=-90]{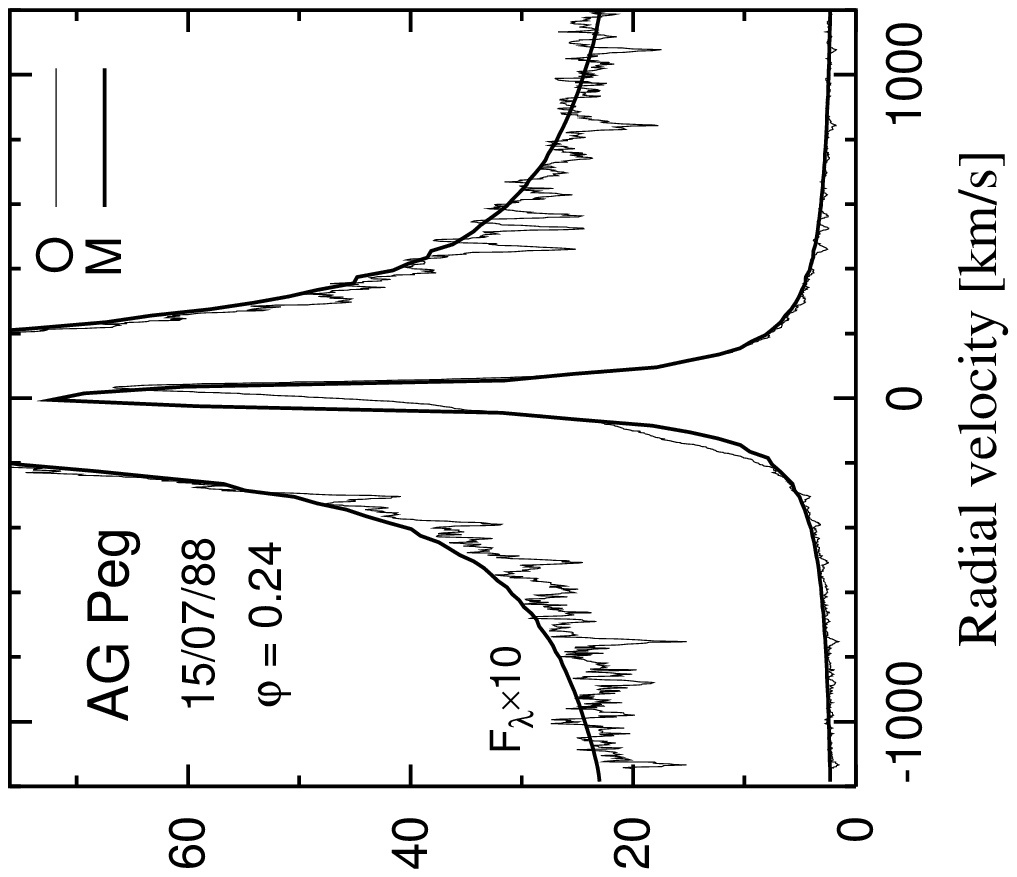}\hspace{5mm}
                      \includegraphics[angle=-90]{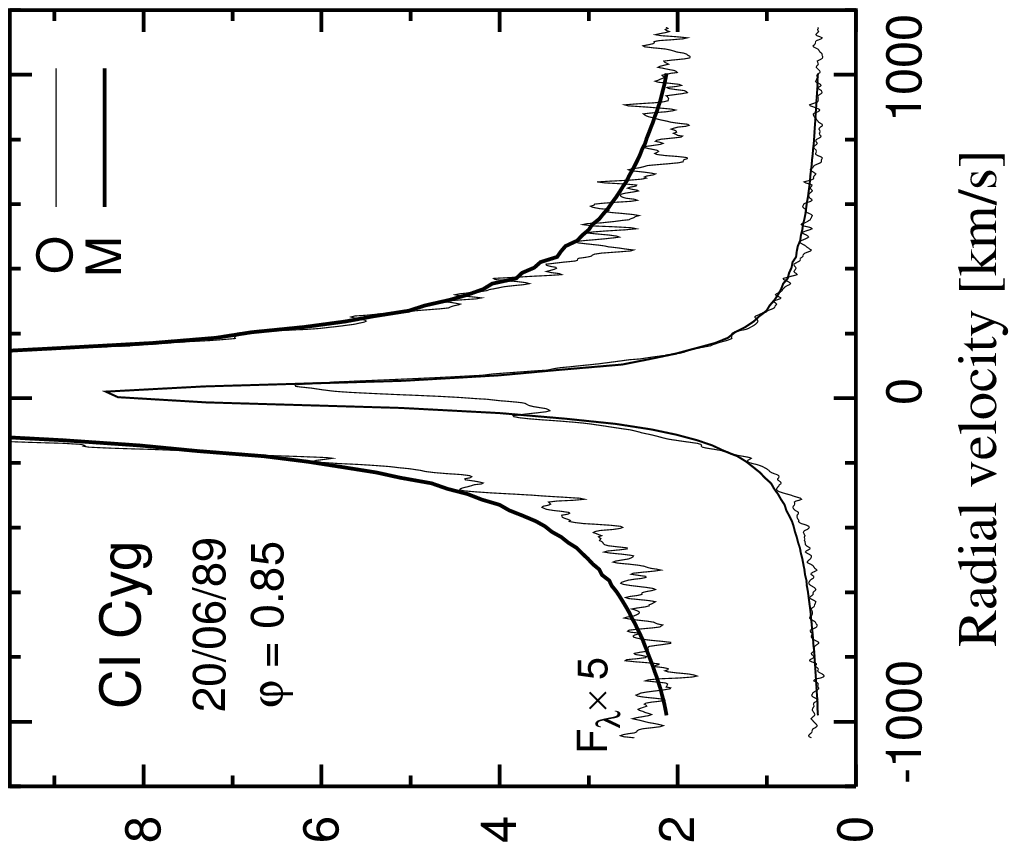}
                      \includegraphics[angle=-90]{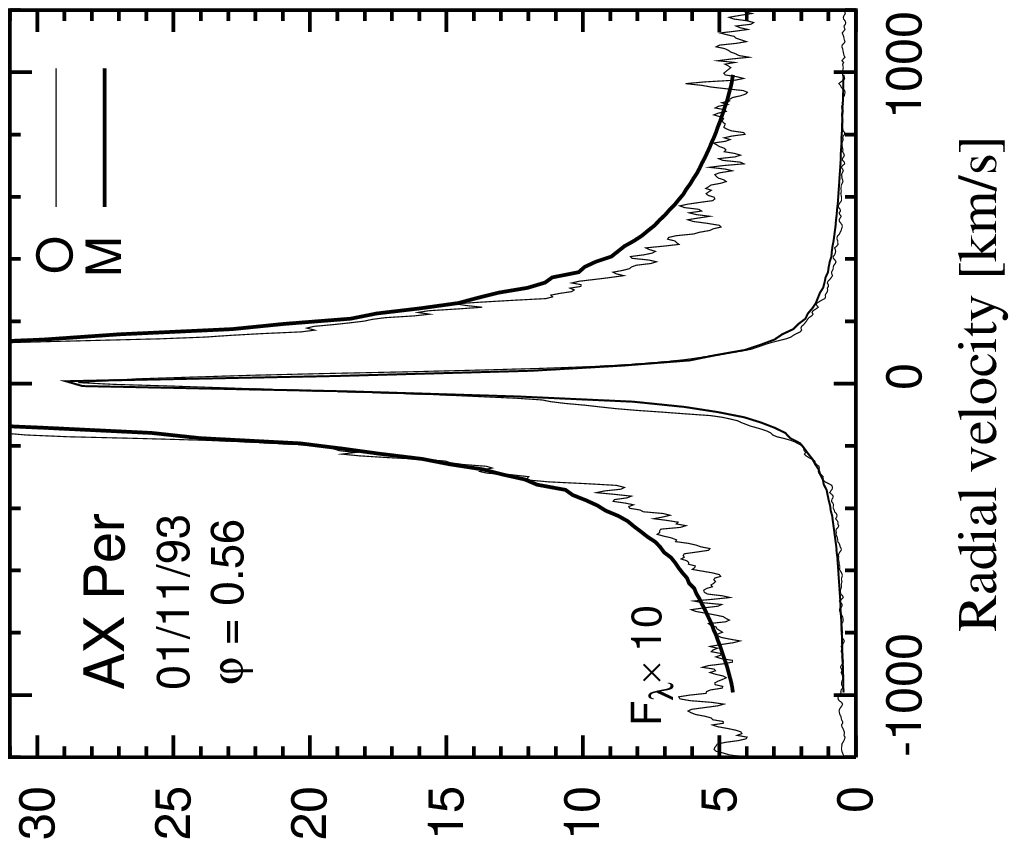}}\vspace{3mm}
%\begin{center}
            {\sf A C T I V E ~ P H A S E S}\\[-1mm]
%\end{center}
\resizebox{\hsize}{!}{\includegraphics[angle=-90]{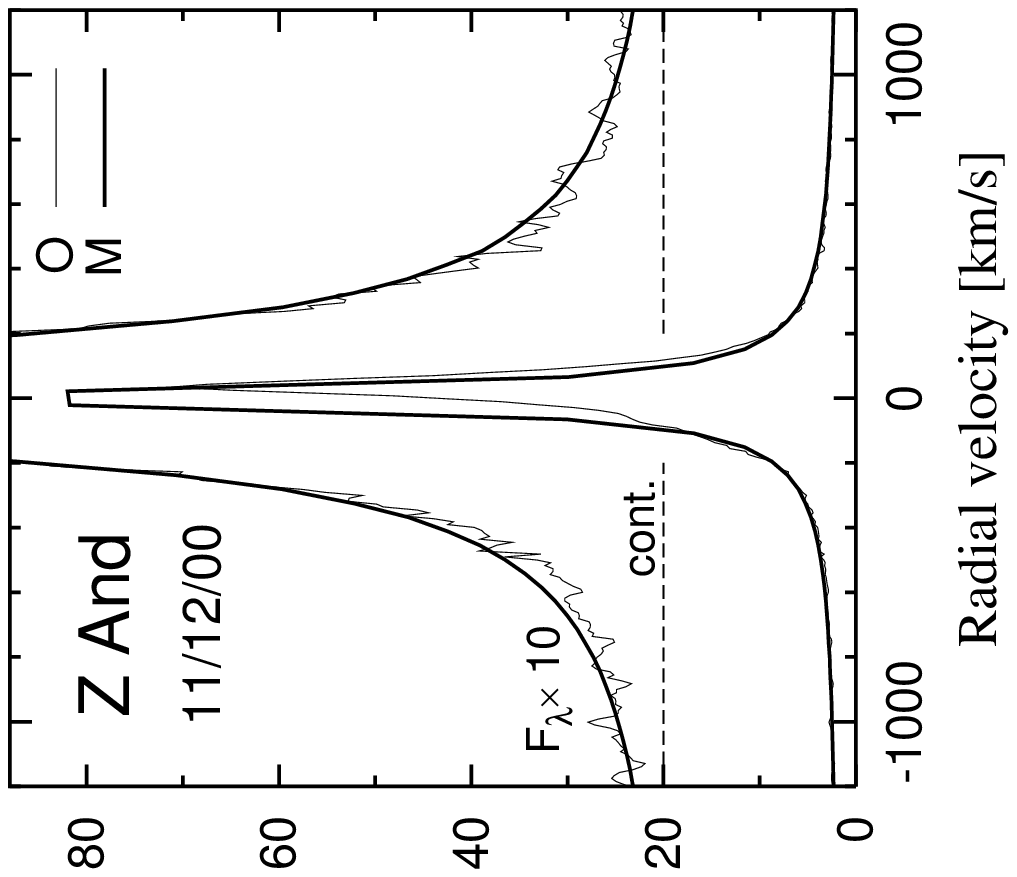}
                      \includegraphics[angle=-90]{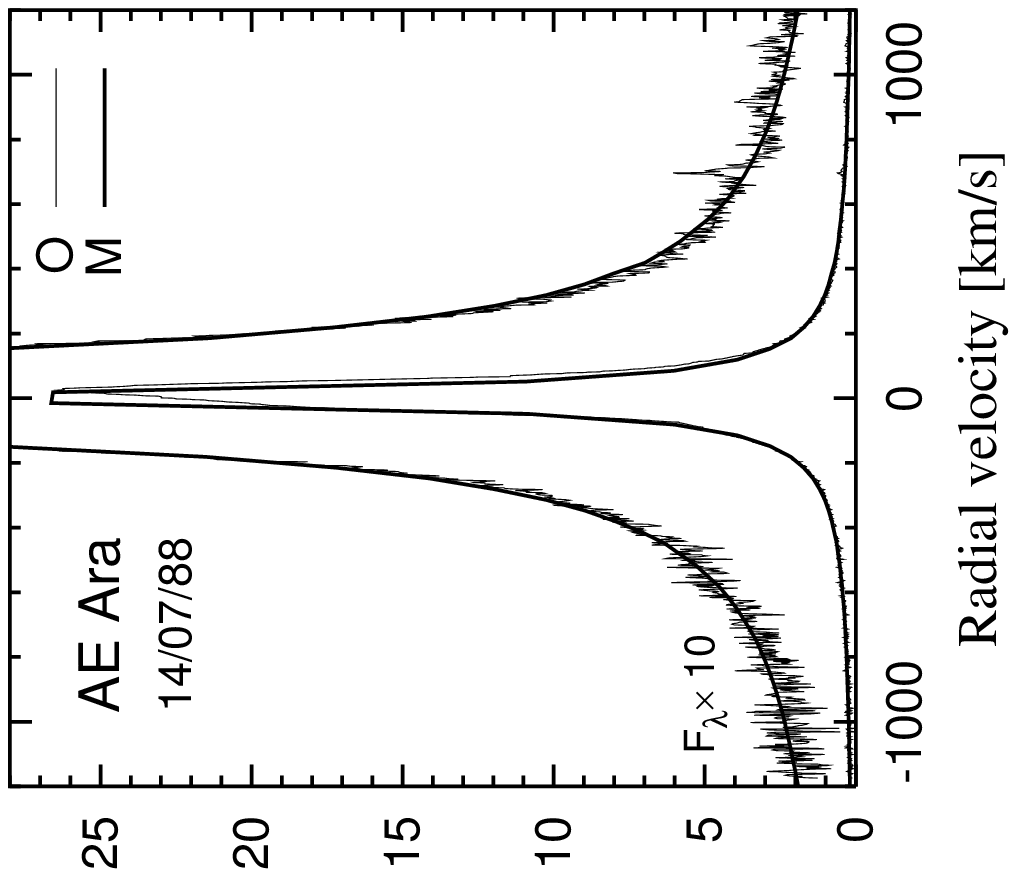}
                      \includegraphics[angle=-90]{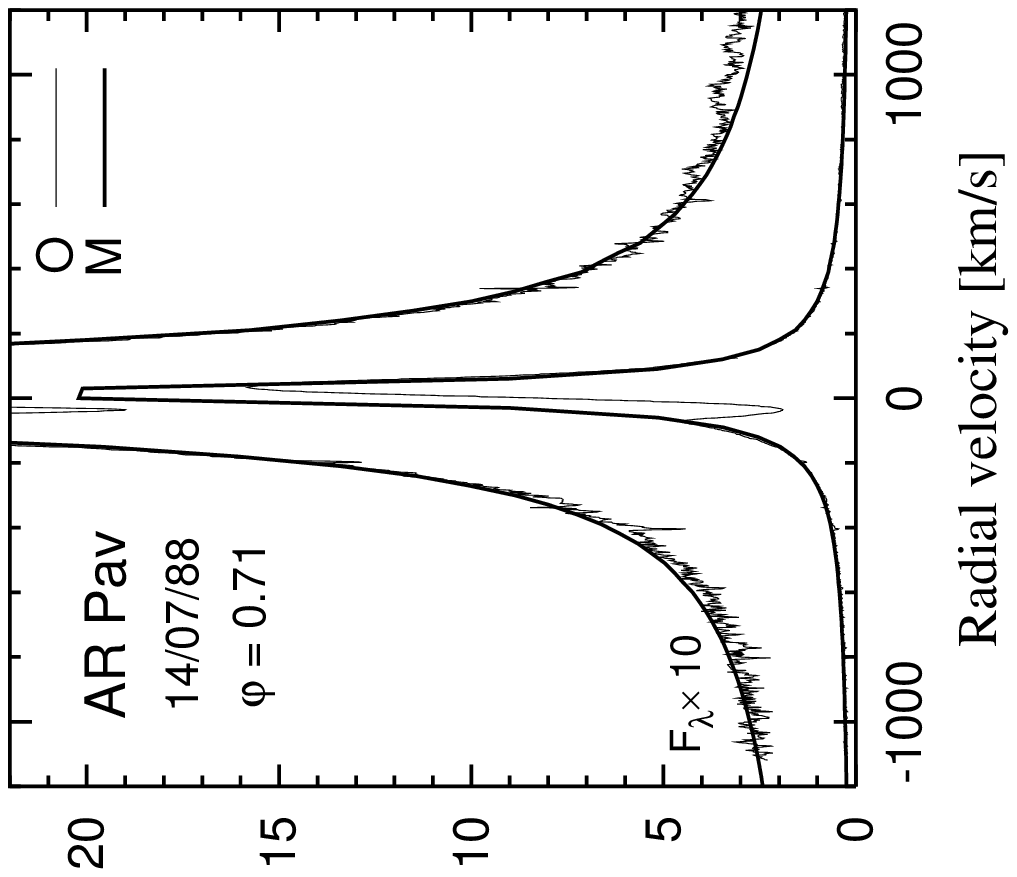}
                      \includegraphics[angle=-90]{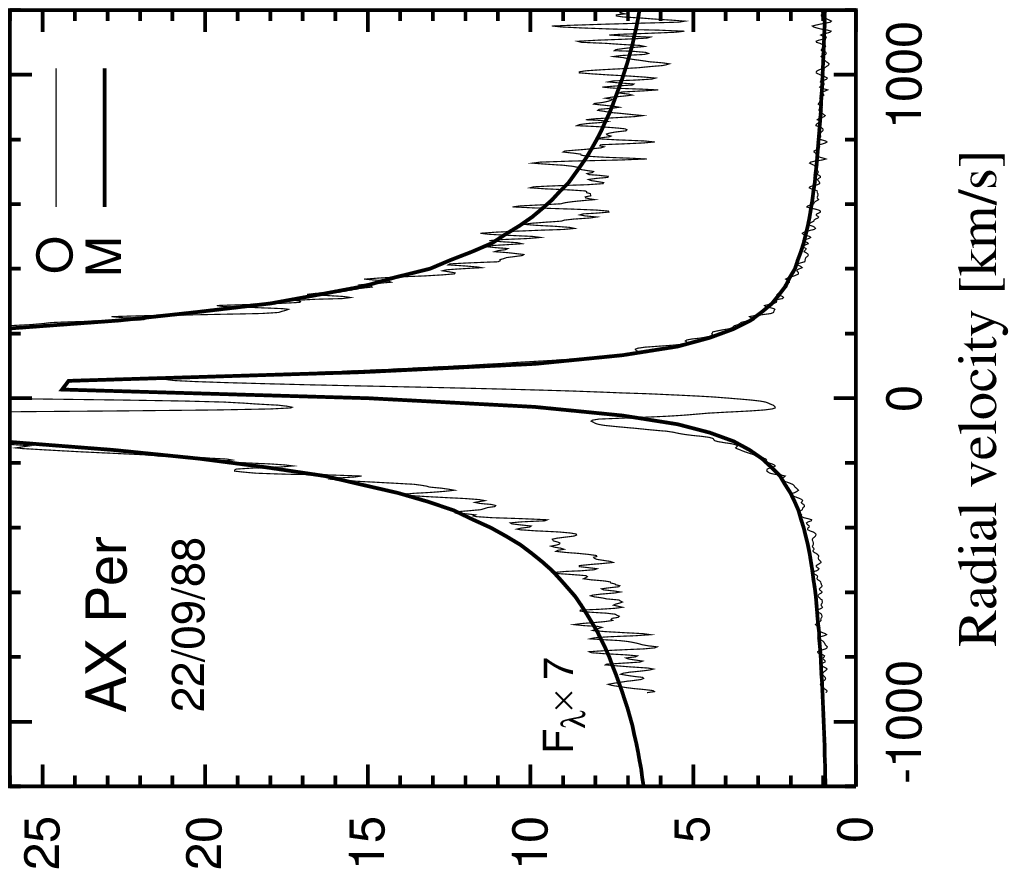}}
\caption[]{
Comparison of the modeled (M) and observed (O) \ha line profiles 
for selected symbiotic stars during their quiescent phases (top 
two raws of panels) and active phases (the bottom raw of panels). 
The systemic velocity was subtracted. The model is described in 
Sect.~2.4 and the corresponding parameters are in Table~1. 
It fits the observed profiles for $|\Delta v| \ga $\,200\kms. 
Fluxes are in $10^{-11}$\ecsa. 
          }
\end{center}
%\label{fig_1}
\end{figure*}

\section{Comparison with observations}

\subsection{Data sources}

In major part we used observations of \ha profiles from the 
survey of \cite{vw+93} and \cite{i+94}. Other sources are 
referred in Table~2. If possible, we selected objects that 
were observed during both quiescence and activity. 
We converted observed fluxes in relative units to fluxes in 
\ecsa\ with the aid of simultaneous optical $V$ and $R$ photometry 
\citep[][ and references therein]{sk+04} and the model 
SED according to \cite{sk05}. 
Approximate corrections for emission lines 
\citep{sk03} were also included. To deredden the fluxes we 
used appropriate $E_{\rm B-V}$ from Table~1 of \cite{sk05}. 
For the purpose of Sect.~4 we also estimated emission measure, 
$EM_{\rm obs}$, of the symbiotic nebula at the dates of \ha 
observations (Table~2). For the sake of simplicity and 
availability we used dereddened fluxes, $F_{\rm U}$, derived 
from $U$-magnitudes. Then according to Eq.~(18) of \cite{sk05}, 
\begin{equation}
 EM_{\rm obs}\,\dot=\,4\pi d^2 \frac{F_{\rm U}}{\varepsilon_{\rm U}}, 
\end{equation}
where $\varepsilon_{\rm U}$ is the volume emission coefficient 
per electron and per ion ($\rm erg\,cm^{3}\,s^{-1}\,\AA^{-1}$). 
We used the average value of $\varepsilon_{\rm U}$ from both 
the sides of the Balmer jump corresponding to the electron 
temperature given by the SED \citep{sk05}. 

\subsection{Model parameters}

Geometrical parameters of our wind model are described in 
Sect.~2.2. Some limits for parameters $R_{\rm D}$ and $H$ 
can be estimated from the effective radius, $R_{\rm h}^{\rm eff}$, 
of the hot star. This parameter represents the radius of 
a sphere that produces the observed luminosity of the hot 
stellar source and can be derived from modeling the SED of 
the ultraviolet continuum \citep{sk05}. In our model 
$4\pi (R_{\rm h}^{\rm eff})^2 = 4\pi R_{\rm D} H$. 
During quiescent phases we observe 
$R_{\rm h}^{\rm eff} \approx 0.15\,R_{\sun}$ \citep[][ Table~3]{sk05}. 
Assuming the ratio $H/R_{\rm D} = 0.1$ then yields 
$R_{\rm D} = 0.5\,R_{\sun}$ and $H = 0.05\,R_{\sun}$. 
During active phases we assume flared disk with $H/R_{\rm D} = 0.3$. 
Then parameters $H$ and $R_{\rm D}$ are adjusted to the 
corresponding $R_{\rm h}^{\rm eff}$ (a few of $R_{\sun}$) 
for objects we investigate here. 

The origin of the wind, $R_{\rm w}$, and $\beta$ in the wind 
law are model parameters, while $v_{\infty}$ is given by 
the extension of the wings. 
Values of $R_{\rm w}$, and $\beta$ are critical for 
the synthetic profile. Generally, a larger value of 
$\beta$ corresponds to a slower and denser wind with 
a higher emissivity at a point $r$. 
Therefore the wind characterized with a larger value of 
$\beta$ produces a narrower profile. Regions close to the 
wind's origin ($R_{\rm w} < H < r \ll r_0$) have the largest 
emissivity, because of high densities for small values of 
both $r$ and $v(r)$; they contribute mainly to the line core. 
However, a fraction of their radiation is blocked by the outer 
rim of the disk in our model. 
In addition, optical properties of these regions can deviate 
from the optically thin case. Therefore we do not aim to fit 
the core of the line by this procedure. At further distances 
($r \ga 1.2\,R_{\rm w}$) the wind is accelerated to 
$v(r) \ga 200 - 250$\kms\ (Eq.~(2), $\beta\sim 1.7$). 
Contributions from these regions are more important to 
create the broad wings. A good agreement between the modeled 
and observed profiles for radial velocities 
$|\Delta v| \ga 200$\kms\ (Fig.~3) is consistent with our 
assumption on the optically thin regime from this distance 
above the source of the wind. 

\subsection{Wing profiles, $\dot M$ and uncertainties}

We reconstructed synthetic profiles according to Eqs.~(10) 
and (12). To obtain an appropriate solution we calculated 
a grid of models for reasonable ranges of 
$R_{\rm w}$ and $\beta$. By this way we also estimated their 
uncertainties to $20\%$ and $10\%$, respectively. A comparison 
of resulting models with observations is shown in Fig.~3 and 
the corresponding parameters are introduced in Table~1. 
Models match perfectly the line profile for $|\Delta v| > 200$\kms. 
Therefore to determine the mass-loss rate from the hot star, 
$\dot M$, we compared the luminosity of wings for 
$|\Delta v| \ga 200$\kms\ ($L_{\alpha}$(200) in our notation) 
with that calculated according to Eq.~(14) for the same radial 
velocity interval. 
Dependences of $L_{\alpha}$(200) on $\dot M$ for our models 
are plotted in Fig.~2. 
Uncertainties in the $\dot M$ values were determined from those 
of $F({\rm H\alpha}$), $\beta$ and $R_{\rm w}$. We estimated 
the uncertainty in the \ha fluxes to $\sim\,10$\%, which results 
mainly from determination of the level of the local continuum 
(Sect.~3.1). As $\dot M \propto \sqrt{L(H\alpha)}$, the uncertainty 
in fluxes increases that of $\dot M$ with only square root. 
We estimated them to 8\% -- 12\%. 

\section{Discussion}

We showed that the profiles from the ionized optically thin, 
bipolar stellar wind match well the observed \ha wings for 
$|\Delta v| \ga 200$\kms. 
The curve fitting the profile is proportional to $\Delta v^{-2}$ 
(Eqs.~10 and 12), which is of the same type as that including 
solely the Raman Ly$\beta \rightarrow {\rm H}\alpha$ scattering 
process \citep{l00}. From this point of view it is not possible 
to distinguish contributions from the ionized wind and the Raman 
scattering in the wing profile directly. However, both the 
processes take place in very different regions of the binary. 
The {\em ionized} stellar wind in our model is located around 
the hot star, while the Raman-scattered photons originate in 
the {\em neutral} part of the wind from the giant. 
Below we summarize observational characteristics of the \ha 
emission that could help to identify main sources of radiation 
contributing to its broad wings. 

\subsection{Observed properties of \ha wings}

(i) 
During eclipses of the hot component by the giant the \ha emission 
in both the core and the wings decreased significantly. 
Examples here are 
  AX\,Per \citep[][ Fig.~6]{sk+01}, 
  AR\,Pav \citep[][ Fig.~2]{q+02}, 
  FN\,Sgr \citep[][ Table~2, Fig.~4]{brandi} and 
  Z\,And \citep[][ Fig.~4]{sk+06}. 
These observations suggest that a significant fraction of the 
broad \ha wings is formed nearby the hot star. 

(ii) 
\cite{it00}, \cite{q+02}, \cite{mika+03} and \cite{brandi} 
revealed that the radial velocities from the wings of \ha follow 
the orbital motion of the hot component in V1329\,Cyg, AR\,Pav, 
AE\,Ara and FN\,Sgr, respectively. This implies that the region 
of the \ha wings formation is connected with the hot star. 

(iii)
The wing profiles from our sample of objects are symmetrically 
placed around $\lambda_0$. 
In cases of CI\,Cyg (20/06/89, $\varphi$ = 0.85) and AR\,Pav 
(17/07/88, $\varphi$ = 0.71) we shifted the model by +12 and 
+15\kms, respectively, to match better observations. These 
shifts are consistent with the hot component orbital motion. 
Only in the case of Z\,And (22/09/88, $\varphi$ = 0.30) 
the model was shifted by +10\kms, which is against the orbital 
motion. Nature of this difference is not clear. 

(iv)
In our sample of \ha profiles we did not find {\em systematic} 
shifts, which could be associated with the Raman scattering. 
For example, the supposed blue-shifted component from 
the neutral wind at/around the binary axis that moves against 
the incident Ly$\beta$ photons 
\citep[][]{schmid+99} and/or the redshift of the \ha wing 
center as suggested by \cite{jl04}, were not indicated. 
From this point of view, a good agreement between systemic 
velocities determined independently from M-giant and 
H$\alpha$-wings radial velocities, respectively, for objects 
referred in the point (ii), is consistent with our finding. 
%
%============================================|
%----- Fig. 4 L(Hw) as a function of EM -----|
%============================================|
%
\begin{figure}
\centering
\begin{center}
\resizebox{\hsize}{!}{\includegraphics[angle=-90]{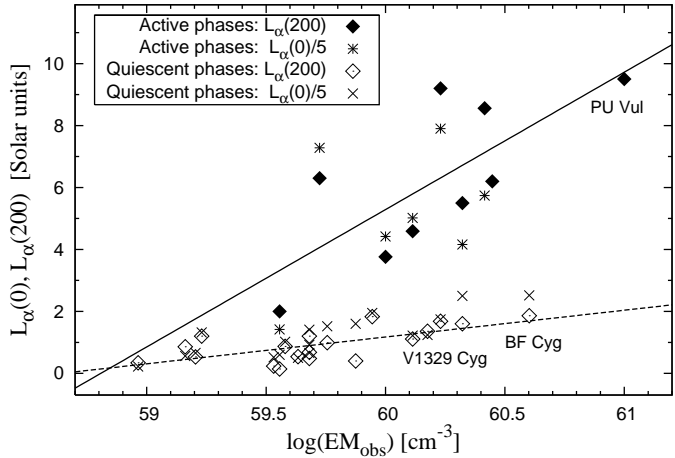}}
\caption[]{
Total \ha line and wing luminosities, $L_{\alpha}$(0) and 
$L_{\alpha}$(200), respectively, as a function of the observed 
emission measure $EM_{\rm obs}$. The total line luminosities 
are divided by a factor of 5 for a better visual display. 
Corresponding quantities are from Table~2. 
          }
\end{center}
%\label{fig_1}
\end{figure}
%
%
%==================================================|
%--------- Table 2: EM and L(H-alpha;200) ---------|
%==================================================|
%
\begin{table}
\caption[]{Emission measures and \ha luminosities}
\begin{center}
\begin{tabular}{cccccc}
\hline
\hline
Object & Date &  Ref. &$EM_{\rm obs}$& $L_{\alpha}$(0)& $L_{\alpha}$(200) \\
       &      &       & [cm$^{-3}$]  &  [$L_{\sun}$]  &    [$L_{\sun}$]   \\
\hline
   \multicolumn{6}{c}{Quiescent phases}             \\
Z\,And     & 30/06/86 & 1,2 & 8.8E+59  &  9.8 & 1.83\\ 
           & 07/07/87 & 1,2 & 5.7E+59  &  7.6 & 0.99\\ 
           & 22/09/88 & 1,2 & 3.8E+59  &  5.1 & 0.87\\ 
AX\,Per    & 01/11/93 &  3  & 1.7E+59  &  6.6 & 1.20\\
RW\,Hya    & 10/07/92 & 1,2 & 9.2E+58  &  1.1 & 0.34\\ 
SY\,Mus    & 14/07/88 & 1,2 & 3.6E+59  &  3.0 & 0.14\\ 
CI\,Cyg    & 21/09/88 & 1,2 & 7.5E+59  &  8.0 & 0.40\\ 
           & 20/06/89 & 1,2 & 4.8E+59  &  3.3 & 0.48\\ 
V1329\,Cyg & 21/09/88 & 1,2 & 2.1E+60  & 12.5 & 1.60\\ 
           & 16/09/89 & 1,2 & 4.8E+59  &  7.1 & 1.20\\
           & 30/06/86 & 1,2 & 1.5E+60  &  6.2 & 1.36\\
           & 06/05/90 & 1,2 & 1.3E+60  &  6.1 & 1.11\\
           & 29/07/91 & 1,2 & 1.7E+60  &  8.7 & 1.68\\
BF\,Cyg    & 20/08/94 & 1,2 & 4.0E+60  & 12.6 & 1.86\\
V443\,Her  & 20/06/89 & 1,2 & 1.6E+59  &  3.3 & 0.53\\
           & 04/05/90 & 1,2 & 3.4E+59  &  2.6 & 0.23\\
AG\,Peg    & 15/07/88 & 1,2 & 4.8E+59  &  4.6 & 0.68\\
           & 07/07/87 & 1,2 & 4.3E+59  &  2.5 & 0.54\\
AG\,Dra    & 18/06/86 & 1,2 & 1.4E+59  &  2.8 & 0.86\\
   \multicolumn{6}{c}{Active phases}                \\
Z\,And     & 11/12/00 &  4  & 2.1E+60  & 20.8 & 5.50\\
           & 02/02/03 &  4  & 2.6E+60  & 28.7 & 8.56\\
           & 31/07/03 &  4  & 1.3E+60  & 25.1 & 4.59\\
           & 13/11/03 &  4  & 1.0E+60  & 22.1 & 3.76\\
AR\,Pav    & 14/07/88 & 1,2 & 1.7E+60  & 39.5 & 9.20\\
AE\,Ara    & 14/07/88 & 1,2 & 5.3E+59  & 36.4 & 6.30\\
BF\,Cyg    & 04/06/93 &  5  & 2.8E+60  & 62.7 & 6.20\\
           & 21/09/88 & 1,2 & 4.1E+60  & 33.8 & 6.10\\
AX\,Per    & 22/09/88 & 1,2 & 3.6E+59  &  7.1 & 2.00\\ 
PU\,Vul    & 28/09/88 & 1,2 & 1.0E+61  & 144  & 9.50$^\dagger$\\
\hline                                              
\end{tabular}
\end{center}
Ref.: 1 -- \cite{vw+93}, 2 -- \cite{i+94},
      3 -- \cite{sk+01}, 4 -- \cite{sk+06}, 
      5 -- \cite{sk+97}\\
$^\dagger$ = $L_{\alpha}$(500)
\end{table}

\subsection{\ha luminosity as a function of the nebular emission}

Here we investigate a relationship between the \ha luminosities, 
$L_{\alpha}$(0), $L_{\alpha}$(200), and the emission measure of
the symbiotic nebula, $EM_{\rm obs}$, which is due to 
photoionization. Relevant data are described in Sect.~3.1, 
summarized in Table~2 and plotted in Fig.~4. 

During {\em quiescent} phases the nebular emission originates 
in the ionized part of the wind from the giant as given by 
the STB \citep{stb} model. This was independently supported 
by finding of \cite{n+88} that symbiotic objects fit well the 
CNO abundance ratios of normal red giants. Following analyses 
showed that the model is applicable for most of quiescent 
symbiotics \cite[e.g.][]{skt93,mio02}. Within this model the 
ionized wind from the giant can contribute to only the \ha core 
emission, because of its small radial velocity dispersion. 
The broad \ha wings thus have to be of a different nature. 
Figure~4 plots the relevant quantities as a function of 
$\log(EM_{\rm obs})$. The observed dependencies are in qualitative 
agreement with the model: The total \ha emission, $L_{\alpha}(0)$, 
is a strong function of $EM_{\rm obs}$, whereas the wing emission, 
$L_{\alpha}$(200), shows only a faint dependence. 
Assuming that the wing emission originates in the hot stellar 
wind, the corresponding mass-loss rates are of a few 
$\times 10^{-8}$\myr\ (Table~1, Fig.~2). However, such the wind 
can produce only very small nebular emission. According to 
Eqs.~(3) and (5) its emission measure, 
$EM_{\rm w} = \int_{V}n_{\rm e}n^{+}[1-w(r)]\,{\rm d}V$,
can be expressed as
\begin{equation}
  EM_{\rm w} = \frac{I_1 + I_2}{4\pi(\mu m_{\rm H})^2}\,
      \Big(\frac{\dot M}{v_{\infty}}\Big)^{2}\frac{1}{R_{\rm w}}.
\end{equation}
Model parameters (Sect.~3.2 and Table~1) yield 
$EM_{\rm w} \approx 10^{58}$\,\cmt. 
Thus during quiescent phases 
\begin{equation}
    EM_{\rm w} \ll EM_{\rm obs},
\end{equation}
because $EM_{\rm obs} \approx 10^{59}$\,\cmt\ (Table~2). 
Qualitatively, larger hot star luminosity gives rise to larger 
emission measure \citep[][ Table~3]{sk05} and probably drives 
a stronger hot star wind. This would explain the 
$L_{\alpha}$(200)/$\log(EM_{\rm obs})$ dependence. 
As $EM_{\rm w} \ll EM_{\rm obs}$, this dependence is faint 
and the wing emission cannot rival that from the ionized giant 
wind. Consequently, the radio emission satisfies that from 
the giant's wind (i.e. the STB model as mentioned above) without 
a detectable influence of the emission from the hot star wind. 

During {\em active} phases the wing emission increases with 
$\log(EM_{\rm obs})$ by a factor of $\sim4.4$ faster than 
during quiescence (Fig.~4). This supports the idea that the 
source of emission producing the broad \ha wings is the ionized 
hydrogen. Our H$\alpha$-wing models then imply that its kinematics 
corresponds to a fast wind from the hot star at rates of 
   a few $\times\,(10^{-7} - 10^{-6})$\myr\ (Table~1). 
The wind produces a large emission measure, because it increases 
with $\dot M^2$ (Eq.~16). Our models yield 
$EM_{\rm w} \ga 10^{59}$\,\cmt, which can rival 
the $EM_{\rm obs}$ (Tables~1 and 2). 
Thus during active phases 
\begin{equation}
    EM_{\rm w} \la EM_{\rm obs}.
\end{equation}
Such the large emission influences also the radio spectrum. 
For example, during the active phase of Z\,And the steepness 
of the radio continuum between 1.4 and 15\,GHz was a factor of 
$\sim$2 larger than during quiescence (Fig.~5), which implies 
optically thicker conditions. Therefore the $cm$ radio emission 
from the activity could be attributed to optically thick $f-f$ 
emission from the hot star wind \citep[cf.][]{st92}. 
However, we need more radio observations for active objects 
and to elaborate corresponding quantitative model taking into 
account their configuration, for example, that suggested by 
\cite[][ Fig.~27]{sk05}. 

Finally, we note that the presence of a strong hot star wind 
during active phases is consistent with the finding of \cite{sk05} 
that the low-temperature nebula (LTN) in active symbiotics has 
a high emissivity ($N_{\rm e} \sim $ a few times $10^{9}$\cmt) 
and is located around the hot star, because it is subject to 
eclipses. Therefore the LTN in active symbiotics may be 
attributed to the emission of the wind from the hot component. 
%
%===============================================|
%----- Fig. 5 Radio observations for Z And -----|
%===============================================|
%
\begin{figure}
\centering
\begin{center}
\resizebox{\hsize}{!}{\includegraphics[angle=-90]{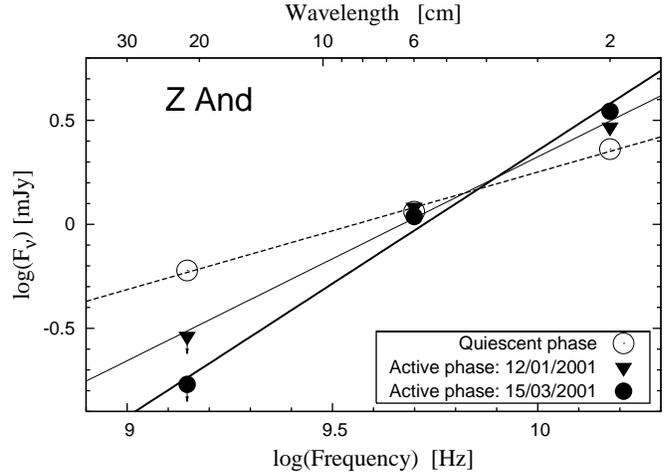}}
\caption[]{
Radio observations of Z\,And at $cm$ wavelengths during 
quiescence and the recent activity. Steeper continuum 
spectrum during activity suggests optically thicker 
conditions than during quiescence. 
The data are from \cite{b+04}. 
          }
\end{center}
%\label{fig_1}
\end{figure}

\section{Concluding remarks}

In this paper we introduced a model of the optically thin, 
bipolar stellar wind from the hot components in symbiotic binaries. 
We derived an expression, which relates the \ha luminosity 
from the wind to the mass-loss rate (Eq.~5) and calculated 
the corresponding line profile with the aid of Eq.~(10). 
We applied the model to the observed \ha profiles for 10 
symbiotic stars during their 
quiescent and active phases. Synthetic profiles provide a good 
fit to wings for $|\Delta v| \ga 200$\kms\ from the line center. 
According to Eq.~(10) the wing profile can be approximated by 
the curve $f(\Delta v) \propto \Delta v^{-2}$, which is of
the same type as that resulting from the Raman scattering 
process \citep{l00}. Therefore it is not possible to distinguish 
between contributions from the ionized hot stellar wind and 
that from the Raman scattering process by only modeling the line 
profile. To support the former possibility we investigated 
relationship between the emission from the \ha wings and 
the emission measure of the symbiotic nebula. 

We found that during quiescent phases the dependence 
$L_{\alpha}$(200)/$\log(EM_{\rm obs})$ is faint (cf. Fig.~4). 
The wing emission is relatively very small (Eq.~17). 
If the wings originate from a fast hot star wind, 
the corresponding mass-loss rates are of a few 
$\times 10^{-8}$\myr\ (Table~1). However, it is difficult to 
indicate their emission independently by other observations. 
Even in the radio the wing emission has no detectable 
effect -- the radio spectra satisfy radiation of the ionized 
wind from the giant (the STB model). 

During active phases the $L_{\alpha}$(200)/$\log(EM_{\rm obs})$ 
relation is a factor of $\sim$\,4.4 steeper than that from 
quiescence. In this case the wing emission can represent 
a significant fraction of the observed nebular emission 
(Eq.~18) and thus can affect the radio spectrum. At $cm$ 
wavelengths the steep continuum spectrum during the recent 
activity of Z\,And (Fig.~5) is consistent with an optically 
thick $f-f$ emission from the hot star wind. 

The $L_{\alpha}$(200)/$\log(EM_{\rm obs})$ relationship and 
other characteristics of the \ha profiles (Sect.~4.1) 
suggest that the ionized hydrogen located around the hot star in 
the form of a fast stellar wind is the dominant source of 
the emission in the $\pm(200\div$2000)\kms\ broad \ha wings 
during active phases. The corresponding mass-loss rates are of 
  a few $\times\,(10^{-7} - 10^{-6})$\myr\ (Table~1).
This finding allow us to attribute the LTN emission in active 
symbiotics to that from the hot star wind. 

\begin{acknowledgements}
The author thanks the anonymous referee for inspiring comments.
This research was in part supported by a grant of the Slovak 
Academy of Sciences No.~2/4014/04. 
\end{acknowledgements}
%
%%%%%%%%%%%%%%%%%%%%%%%%%%%%%%%%%%%%%%%%%%%%%%%%%%%%%%%%%%%%%%%%%
%


\begin{thebibliography}{}
\bibliographystyle{aa}

\bibitem[Arrieta \& Torres-Peimbert (2003)]{at03}
         Arrieta, A., \& Torres-Peimbert, S. 
         2003, ApJS, 147, 97

\bibitem[Bertout et al. (1985)]{b+85}
         Bertout, C., Leitherer, C., Stahl, O., \& Wolf, B. 
         1985, A\&A, 144, 87

\bibitem[Brandi et al. (2005)]{brandi}
         Brandi, E., Mikolajewska, J., Quiroga, C., et al. 
         2005, A\&A, 440, 239

\bibitem[Brocksopp et al. (2004)]{b+04}
         Brocksopp, C., Sokoloski, J. L., Kaiser, C., et al. 
%Richards, A.M., Muxlow, T.W.B., \& Seymour, N. 
         2004, MNRAS, 347, 430

\bibitem[Castor et al. (1975)]{cak}
         Castor, J. I., Abbott, D. C., \& Klein R. I. 
         1975, ApJ, 195, 157

\bibitem[Chochol et al. (2005)]{chochol+05}
         Chochol, D., Katysheva, N. A., Pribulla, T., et al. 
         2005, Contrib. Astron. Obs. Skalnat\'e Pleso, 35, 107

\bibitem[Crocker et al. (2001)]{crok+01}
         Crocker, M. M., Davis R. J., Eyres S. P. S., et al. 
%Bode M.F., Taylor A.R., Skopal A., Kenny H.T., 
         2001, MNRAS, 326, 781

\bibitem[Crocker et al. (2002)]{crok+02}
         Crocker, M. M., Davis R. J., Spencer, R. E., et al. 
%Eyres S. P. S., et al. Bode M. F., Skopal A., 
         2002, MNRAS, 335, 1100

\bibitem[Eriksson et al. (2004)]{ejw}
         Eriksson, M., Johansson, S., \& Wahlgren, G. M. 
         2004, A\&A, 422, 987

\bibitem[Eyres et al. (2005)]{eyres+05}
         Eyres, S. P. S., Heywood, I., O'Brien, T. J., et al. 
         2005, MNRAS, 358, 1019

\bibitem[Fern\'andez-Castro et al. (1995)]{fc+95}
         Fern\'andez-Castro, T., Gonz\'alez-Riestra, R.,
         Cassatella, A., Taylor, A. R., \& Seaquist E. R.
         1995, ApJ, 442, 366

\bibitem[Galloway \& Sokoloski (2004)]{g+s04}
         Galloway, D. K., \& Sokoloski, J. L. 
         2004, ApJ, 613, L61

\bibitem[Ikeda \& Tamura (2000)]{it00}
         Ikeda, Y., \& Tamura, S. 2000, PASJ, 52, 589

\bibitem[Ivison et al. (1994)]{i+94}
         Ivison, R. J., Bode, M. F., \& Meaburn, J.
         1994, A\&AS, 103, 201

\bibitem[Jung \& Lee (2004)]{jl04}
         Jung,Y-Ch., \& Lee, H-W. 2004, MNRAS, 350, 580

\bibitem[Kenny et al. (1996)]{kenny}
         Kenny, H. T., Taylor, A. R., Frei, B. D., et al. 
         1996, in Radio Emission from the Stars and the Sun, 
         ed. A. R. Taylor \& J. M. Paredes, ASP Conf. Ser., 
         93, 197

\bibitem[Lamers, \& Cassinelli (1999)]{lc99}
         Lamers, H. J. G. L. M., \& Cassinelli, L. P. 
         1999, Introduction to Stellar Winds, CUP, Cambridge

\bibitem[Lee (2000)]{l00}
         Lee, H-W. 2000, ApJ, 541, L25

\bibitem[Lee \& Hyung (2000)]{lh00}
         Lee, H-W., \& Hyung, S. 2000, ApJ, 530, L49

\bibitem[Mikolajewska et al. (2002)]{mio02}
         Mikolajewska, J., Ivison, R. J., \& Omont, A. 2002,
         Adv. Space Res., 30, 2045

\bibitem[Mikolajevska et al. (2003)]{mika+03}
         Mikolajewska, J., Quiroga, C., Brandi, E., et al.
         2003, in Symbiotic Stars Probing Stellar Evolution,
         ed. R. L. M. Corradi, J. Mikolajewska \& T. J. Mahoney,
         ASP Conf. Ser., 303, 147

\bibitem[Nussbaumer et al. (1988)]{n+88}
         Nussbaumer, H., Schild, H., Schmid, H. M., \& Vogel, M.
         1988, A\&A, 198, 179

\bibitem[Nussbaumer et al. (1989)]{nsv89}
         Nussbaumer, H., Schmid, H. M., \& Vogel, M. 
         1989, A\&A, 211, L27

\bibitem[Nussbaumer et al. (1995)]{nsv95}
         Nussbaumer, H., Schmutz, W., \& Vogel, M.
         1995, A\&A, 293, L13

\bibitem[Quiroga et al. (2002)]{q+02}
         Quiroga, C., Mikolajewska, J., Brandi, E., et al. 
% Ferrer, O., \& Garc\'{\i}a L. 
         2002, A\&A, 387, 139

\bibitem[Robinson et al. (1994)]{r+94}
         Robinson, K., Bode, M. F., Skopal, A., et al. 
% Ivison, R. J., \& Meaburn, J. 
         1994, MNRAS, 269, 1

\bibitem[Rudy et al. (1999)]{rudy}
         Rudy, R. J., Meier, S. R., Rossano, G. S., et al. 
         1999, ApJS, 121, 533

\bibitem[Schmid et al. (1999)]{schmid+99}
         Schmid, H. M., Krautter, J., Appenzeller, I., et al.
         1999, A\&A, 348, 950

\bibitem[Schmutz et al. (1994)]{schmutz}
         Schmutz, W., Schild, H., M\"urset, U., \& Schmid, H. M. 
         1994, A\&A, 288, 819

\bibitem[Seaquist, Taylor \& Button (1984)]{stb}
         Seaquist, E. R., Taylor, A. R., \& Button, S. 1984,
         ApJ, 284, 202 (STB)

\bibitem[Seaquist \& Taylor (1992)]{st92}
         Seaquist, E. R., \& Taylor, A. R.
         1992, ApJ, 387, 624

\bibitem[Seaquist et al. (1993)]{skt93}
         Seaquist, E. R., Krogulec, M., \& Taylor, A. R.
         1993, ApJ, 410, 260

\bibitem[Skopal (2003)]{sk03}
         Skopal, A. 2003, Baltic Astron. 12, 604

\bibitem[Skopal (2005)]{sk05}
         Skopal, A. 2005, A\&A, 440, 995

\bibitem[Skopal et al. (1997)]{sk+97}
         Skopal, A., Vittone, A. A., Errico, L., et al. 1997, 
         MNRAS, 292, 703

\bibitem[Skopal et al. (2001)]{sk+01}
         Skopal, A., Teodorani, M., Errico, L., et al. 
         2001, A\&A, 367, 199

\bibitem[Skopal et al. (2002)]{sk+02}
         Skopal, A., Bode, M. F., Crocker, M. M., et al. 
% Eyres, S. P. S., Drech\-sel, H., \& Kom\v{z}\'{\i}k, R. 
         2002, MNRAS, 335, 1109

\bibitem[Skopal et al. (2004)]{sk+04}
         Skopal, A., Pribulla, T., Va\v{n}ko, M., et. al. 
% Veli\v{c}, Z., Semkov, E., Wolf, M., \& Jones, A. 
         2004, Contrib. Astron. Obs. Skalnat\'e Pleso, 34, 45 
         [ArXiv: astro-ph/0402141]

\bibitem[Skopal et al. (2006)]{sk+06}
         Skopal, A., Vittone, A. A., Errico, L., et al. 
         2006, A\&A, 453, 279

\bibitem[Sokoloski et al. (2006)]{sok+06}
         Sokoloski, J. L., Kenyon, S. J., Espey, B. R., et al. 
         2006, ApJ, 636, 1002

\bibitem[Walder \& Folini (2000)]{w+f00}
         Walder, R., \& Folini, D. 2000, in Thermal and Ionization 
         Aspects of Flows from Hot Stars: Observations and Theory, 
         eds. H. J. G. L. M. Lamers, and A. Sapar, 
         ASP Conf. Ser. 204, 331

\bibitem[Wallerstein et al. (1984)]{w+84}
         Wallerstein, G., Willson, L. A., Salzer, J., \& Brugel, E.
         1984, A\&A, 133, 137

\bibitem[van Winckel et al. (1993)]{vw+93}
         van Winckel, H., Duerbeck, H. W., \& Schwarz, H. E. 
         1993, A\&AS, 102, 401
%
\end{thebibliography}
\end{document}